\documentclass[a4paper,11pt]{article}

\usepackage[latin1]{inputenc}
\usepackage[top=2.5cm, left=2.5cm, right=2cm, bottom=2.5cm]{geometry}
\usepackage{graphicx}
\usepackage[centertags]{amsmath}
\usepackage{amsfonts}
\usepackage{amssymb}
\usepackage{amsthm}
\usepackage{newlfont}

\newlength{\defbaselineskip}
\newcommand{\setlinespacing}[1]%
           {\setlength{\baselineskip}{#1 \defbaselineskip}}

\begin{document}

\begin{center}

{\Large Non-Involutive Constrained Systems and Hamilton-Jacobi Formalism}

\vspace{1cm}

M. C. Bertin\footnote{mcbertin@ift.unesp.br}, B. M. Pimentel\footnote{pimentel@ift.unesp.br}, C. E. Valcárcel\footnote{valcarcel@ift.unesp.br}

\vspace{.5cm}

\emph{Instituto de Física Teórica - São Paulo State University},

\emph{Rua Pamplona 145, 01405-900, São Paulo, SP, Brazil}.

\end{center}

\vspace{.25cm}

\begin{abstract}

In this work we discuss the natural appearance of the Generalized Brackets in systems with non-involutive (equivalent to second class) constraints in the Hamilton-Jacobi formalism. We show how a consistent geometric interpretation of the integrability conditions leads to the reduction of degrees of freedom of these systems and, as consequence, naturally defines a dynamics in a reduced phase space.

\vspace{.5cm}


\end{abstract}

\section{Introduction}

Constraints in Lagrangian systems appear when the Hessian matrix built from the Lagrangian function is singular. Since the Hessian is the Jacobian of the Legendre transformation that takes velocities to conjugate momenta, the non existence of an inverse transformation leads to a phase space that is not isomorphic to the tangent bundle of the configuration space. In this phase space the symplectic structure is degenerate and a unique Hamiltonian cannot be defined, leading to problems in the usual construction of the Hamiltonian formalism. This problem was first treated by Dirac's \cite{dirac, dirac1, dirac2} canonical approach, in which the constraints are added to the canonical Hamiltonian function with Lagrange multipliers. The so called Dirac's algorithm is based on consistency conditions, that imposes the constraints as invariants over time evolution.

Other alternative approaches have been formulated since, including the Hamilton-Jacobi (HJ) approach, first developed in \cite{guler} and improved later in \cite{hj1, hj2, hj3}, based on Carathéodory's equivalent Lagrangian method \cite{cara}. The Carathéodory's approach to HJ formalism has a very profound geometrical meaning: the necessary and sufficient condition for an Action to be minimized in a region of the configuration space is the existence of a family of surfaces everywhere orthogonal to a congruence of curves in this space. The family is a solution, in the case of regular systems, of a single HJ partial differential equation (HJPDE). If the system is singular with $k$ constraints we will not have one, but a set of $k+1$ HJPDE. The solution of the characteristics equations related to the HJPDE is the congruence of curves, called simply the characteristics, which forms a dynamical system with several independent variables. This geometrical interpretation is called in the literature  \cite{cara} ``the complete figure'' of variational calculus.

The integrability conditions are in the central role of the HJ formalism for singular systems. They should guarantee, supposing linear independence of the independent variables, the existence of a complete set of HJPDE and, by consequence, the integrability of the characteristics equations. The Frobenius' integrability condition sets up that the constraints must form a complete system in involution with the Poisson Brackets (PB). However, there are systems that do not obey this condition, presenting non-involutive constraints. To circumvent this problem some authors elaborate alternative methods \cite{guler1, rothe}, out of the scope of the HJ formalism.

We will show that imposing the Frobenius' conditions in the form $d\{\phi\} = 0$, in which $\{\phi\}$ represents the set of all constraints, we can deal with non-involutive constraints in a very natural way, allowing linear dependence of the former independent variables. The integrability conditions in this form also leads naturally to a non degenerate symplectic structure in a reduced phase space, represented by some Generalized Brackets (GB). The appearance of GB in the HJ formalism was already noticed in first order Lagrangians \cite{bertin}, but in this work we intend to make a more general formulation, to be valid for general, not only first order systems.

\section{The Complete Figure}

Let it be a Lagrangian function $L(y,\dot{y},x,\dot{x},t)$, with $n$ degrees of freedom, whose Hessian matrix is singular of rank $m$. This means that there are $k=n-m$ constraints $\phi_z \equiv \pi_z - \partial L / \partial \dot{x}^z=0$ and the system has its configuration space separated in two sub-spaces, namely the space of $m$ variables $y$, relate to the invertible part of the Hessian, and the space of $k$ variables $x$, related to the non invertible part. The equation
\begin{equation*}
     \pi_0 + p_a\dot{y}^a + \pi_z\dot{x}^z - L = 0\ ,
\end{equation*}
where $\pi_0 \equiv \partial_tS$, $p_a = \partial_aS$ and $\pi_z = \partial_zS$ and the indexes are given by $\{a\} = \{1, \ldots, m\}$, $\{z\} = \{1, \ldots, k\}$ and $\{i\} = \{1, \ldots, n\}$, is the sufficient condition for an extremum of the action $\int L dt$. Wherever the constraints are valid this equation becomes a HJPDE with the canonical Hamiltonian defined by
\begin{equation}\label{01}
    H_0 \equiv p_a\dot{y}^a + \pi_z\dot{x}^z - L\ ,
\end{equation}
which does not depend on the undetermined velocities $\dot{x}$. Hence, the system must obey the set of $k+1$ HJPDE
\begin{equation}\label{02}
    \phi_\alpha \equiv X_\alpha(S) = \pi_\alpha + H_\alpha = 0\ ,\ \ \ \ \ \ \ \ \ \ \{\alpha\} = \{0,1,\ldots,k\}\ ,
\end{equation}
in which $X_\alpha=\chi^i_\alpha \partial_i$ is a set of vector fields, $\chi^z_\alpha = \delta^z_\alpha$ and $\chi^a_\alpha = \partial \phi_\alpha / \partial p_a$. In our notation $x^0$ is related to the time parameter and $H_z \equiv - \partial L / \partial \dot{x}^z$.

The characteristic equations are given by
\begin{eqnarray}
  &&dy^a = \chi^a_\alpha dx^\alpha = \left\{y^a, \phi_\alpha\right\}dx^\alpha\ , \\
  &&dp_a = \left\{p_a, \phi_\alpha\right\}dx^\alpha\ , \\
  \label{05}
  &&dS = p_ady^a - H_\alpha dx^\alpha\ ,
\end{eqnarray}
where we used Poisson brackets. With equation (\ref{05}) we are able, if desirable, to build an explicit solution for the HJPDE, which gives the family of surfaces orthogonal to the characteristic curves.

\section{Integrability}

The vector fields $X_\alpha$ form a basis of vectors in the tangent space of the family of surfaces, as it can easily be seen by the HJPDE $X_\alpha(S)=\chi^i_\alpha \partial_iS = \chi^i_\alpha p_i=0$. For integrability we need that these vectors form a complete set of linear independent (LI) Hamiltonian vector fields. The necessary and sufficient condition is given by the Frobenius' integrability condition: the Lie derivative of a member of the base with respect to the Hamiltonian flow generated by other member must be a vector in this tangent space. In other words, they must close a Lie algebra:
\begin{equation}\label{06}
    \left[X_\alpha, X_\beta\right] = C_{\alpha\beta}^{\ \ \ \gamma}X_\gamma\ .
\end{equation}
Since the constraints are the generators of the flows, this condition is reflected on the system of HJPDE which must form a system in involution:
\begin{equation}\label{07}
    \left\{\phi_\alpha, \phi_\beta\right\} = C_{\alpha\beta}^{\ \ \ \gamma}\phi_\gamma\ .
\end{equation}

However, systems in physics do not, in general, respect this condition only with the constraints extracted from the Lagrangian. The problem rests in two situations: it may happen that the set of HJPDE is not complete. In this case the integrability conditions should provide new constraints to complete the system. This scenario is actually well supported by the conditions in the form (\ref{07}). Of course, if the final complete set of constraints closes a Lie algebra the system is completely integrable.

The other situation, which is not covered by condition (\ref{07}), is the case in which the vectors in the tangent space of the family are not really LI. To deal with this case we will use as integrability conditions the equations
\begin{equation}\label{08}
    d\phi_\alpha = C_{\alpha\beta}^{\ \ \ \gamma}\phi_\gamma = 0\ ,
\end{equation}
with use of the constraints equations. However, it must be noticed that (\ref{07}) is covered, since $d\phi_\alpha = \left\{\phi_\alpha, \phi_\beta\right\}dx^\beta$. This is important since the Frobenius' condition is a theorem and must be satisfied for every integrable system.

The conditions (\ref{08}) can lead to three situations. It can result in expressions of the type $f(y,x,p,\pi)=0$ and these expressions are the ones we are interested in first place, because they must be considered as constraints in equality to the former set. They must obey the integrability conditions as well and we should be aware of the appearance of other constraints. When all possible constraints are found they must be inserted in the formalism.

However, there will be no independent variables related to the constraints that come from the imposition of integrability. We will deal with this problem by expanding the space of independent variables, relating arbitrary parameters to each new constraint. The set of all constraints, $\phi_z$ in which the index $z$ now cover all the expanded parameter space, is supposed to be complete and the new characteristic equations can be derived from the fundamental differential
\begin{equation}\label{09}
    dF = \left\{F, \phi_\alpha\right\}dx^\alpha\ ,
\end{equation}
when, now, $x^\alpha$ is the set of all independent variables, including the ones related to the new constraints.

At this stage we must use the differential (\ref{09}) to test the integrability of all constraints again. It can happen that a subset, if not all the constraints, identically satisfies the conditions. This is the second situation that will happen when this set is actually in involution, obeying the condition (\ref{07}). If all HJPDE are in this case the system becomes completely integrable. If only a subset is in involution, the system is only partially integrable.

\section{The Generalized Brackets}

The first two possible results of the integrability conditions (\ref{08}), related above, are in agreement with the conditions (\ref{07}). There is, however, a third possibility not predicted by the latter. It may happen that the final set of constraints turns out to be not in involution with the PB.

One way to deal with this problem, using (\ref{07}), is expanding the space of constraints finding new vector fields that closes, finally, a system in involution \cite{guler1}. However, the formalism itself cannot give a consistent way to find these constraints: any system that leaves the final set in involution is a good one. Other way \cite{rothe} is performing a canonical transformation that takes pairs of non-involutive constraints into conjugate variables.

The conditions (\ref{08}) provides a more natural way to deal with non-involutive constraints. It permits linear dependence (LD) of the tangent vector fields $X_\alpha$.

Let us suppose that all constraints are found and form a complete set of HJPDE, $\phi_\alpha=0$, which are related to a set of independent variables $x^\alpha$. Separating the time variable from the other parameters the integrability conditions (\ref{08}) reads
\begin{equation*}
    d\phi_\alpha = \left\{\phi_\alpha, \phi_0\right\}dt + \left\{\phi_\alpha, \phi_z\right\}dx^z = 0\ .
\end{equation*}
First, we analyze the conditions for $\phi_x$:
\begin{equation*}
\left\{\phi_x, \phi_0\right\}dt + \left\{\phi_x, \phi_z\right\}dx^z = 0\ .
\end{equation*}

Let us define an antisymmetric matrix from the PB between the constraints $\phi_z$, $M_{xz} \equiv \left\{\phi_x, \phi_z\right\}$. Then we can write
\begin{equation}\label{10}
M_{xz}dx^z = - \left\{\phi_x, \phi_0\right\}dt\ .
\end{equation}
If the complete system of HJPDE is not in involution, we deal with a regular $M$ matrix. In this case it becomes clear that all vectors $X_\alpha$ are LD from the total differential equations
\begin{equation}\label{11}
dx^z = - (M^{-1})^{zx}\left\{\phi_x, \phi_0\right\}dt\ .
\end{equation}
Notice that the fundamental differential (\ref{09}) becomes
\begin{eqnarray*}
  dF &=& \left\{F, \phi_\alpha\right\}dx^\alpha = \left\{F, \phi_0\right\}dt + \left\{F, \phi_z\right\}dx^z \\
     &=& \left[\left\{F, \phi_0\right\} - \left\{F, \phi_z\right\}(M^{-1})^{zx}\left\{\phi_x, \phi_0\right\}\right]dt\ .
\end{eqnarray*}

Therefore, we redefine the dynamics by eliminating all the independent variables with exception of $t$. It allows us to define the Generalized Brackets
\begin{equation}\label{12}
    \left\{F,G\right\}^* \equiv \left\{F,G\right\} - \left\{F, \phi_z\right\}(M^{-1})^{zx}\left\{\phi_x, G\right\}\ ,
\end{equation}
which have all the properties of the PB: it is a bilinear antisymmetric operator that obeys the Jacobi identity and the Leibniz rule. With the GB the dynamics is given by
\begin{equation}\label{13}
    dF = \left\{F, \phi_0\right\}^*dt\ .
\end{equation}
Besides, all the constraints are in involution with the GB, $\left\{\phi_\alpha, \phi_\beta\right\}^* = 0$. There is no need to test the integrability of $\phi_0$, we can easily see that it is identically satisfied with the GB.

Let us suppose the case in which the $M$ matrix is singular of rank $r\leq k$. In this case there will be a set of $r$ non-involutive constraints. The equations (\ref{10}) can be written by
\begin{equation*}
M_{x\bar{a}}dx^{\bar{a}} + M_{x\bar{z}}dx^{\bar{z}} = - \left\{\phi_x, \phi_0\right\}dt\ ,
\end{equation*}
where $\{\bar{a}\}=\{1,\ldots,r\}$ and $\{\bar{z}\}=\{r+1,\ldots,k\}$. There are two sets of equations:
\begin{eqnarray}
  \label{14}
  M_{\bar{b}\bar{a}}dx^{\bar{a}} + M_{\bar{b}\bar{z}}dx^{\bar{z}} &=& - \left\{\phi_{\bar{b}}, \phi_0\right\}dt \\
  \label{15}
  M_{\bar{x}\bar{a}}dx^{\bar{a}} + M_{\bar{x}\bar{z}}dx^{\bar{z}} &=& - \left\{\phi_{\bar{x}}, \phi_0\right\}dt \ .
\end{eqnarray}
The first set gives
\begin{equation*}
    M_{\bar{b}\bar{a}}dx^{\bar{a}} = - \left\{\phi_{\bar{b}}, \phi_0\right\}dt - \left\{\phi_{\bar{b}}, \phi_{\bar{z}}\right\}dx^{\bar{z}} =  - \left\{\phi_{\bar{b}}, \phi_{\bar{\alpha}}\right\}dx^{\bar{\alpha}}\ ,
\end{equation*}
making $\{\bar{\alpha}\}=\{0,r+1,\ldots,k\}$. Since the rank of $M_{xz}$ is $r$, $M_{\bar{b}\bar{a}}$ is regular and we can write
\begin{equation}\label{16}
    dx^{\bar{a}}  = - (M^{-1})^{\bar{a}\bar{b}}\left\{\phi_{\bar{b}}, \phi_{\bar{\alpha}}\right\}dx^{\bar{\alpha}}\ .
\end{equation}

Hence, we can eliminate $r$ independent variables. The fundamental differential becomes
\begin{eqnarray}\label{17}
    dF &=& \left\{F, \phi_{\bar{a}}\right\}dx^{\bar{a}} + \left\{F, \phi_{\bar{\alpha}}\right\}dx^{\bar{\alpha}}\nonumber\\
    &=& \left\{F, \phi_{\bar{\alpha}}\right\}dx^{\bar{\alpha}} - \left\{F, \phi_{\bar{a}}\right\}(M^{-1})^{\bar{a}\bar{b}}\left\{\phi_{\bar{b}}, \phi_{\bar{\alpha}}\right\}dx^{\bar{\alpha}}\nonumber\\
    &=& \left\{F, \phi_{\bar{\alpha}}\right\}^*dx^{\bar{\alpha}}\ ,
\end{eqnarray}
in which we have the GB
\begin{equation}\label{18}
    \left\{F,G\right\}^* \equiv \left\{F,G\right\} - \left\{F, \phi_{\bar{a}}\right\}(M^{-1})^{\bar{a}\bar{b}}\left\{\phi_{\bar{b}}, G\right\}\ .
\end{equation}

The set of equations (\ref{15}) gives $\left\{\phi_{\bar{x}}, \phi_{\bar{\alpha}}\right\}^*dx^{\bar{\alpha}} = 0$. If $M_{xz}$ has rank $r$, the structure of the GB gives $\left\{\phi_{\bar{x}}, \phi_{\bar{z}}\right\} = 0$ and $\left\{\phi_{\bar{x}}, \phi_{\bar{a}}\right\}=0$, which left us with the conditions
\begin{equation}\label{19}
    \left\{\phi_{\bar{x}}, \phi_0\right\}^* = 0\ .
\end{equation}

In the beginning of this analysis we made the supposition that all the constraints that could come from integrability conditions are computed in $\phi_\alpha$. In this case the above conditions are just $0=0$, identically satisfied. But we could define the GB for any set of non-involutive constraints. If there is any new constraint yet to be found it will appear in conditions (\ref{19}).

\section{Free particle on a surface}

In this example we will apply the described procedure in detail in order to clarify the main aspects of the HJ formalism. Let us consider a free particle in an Euclidian space $\mathbb{R}^n$ restricted in a $m$ dimensional surface $\mathcal{M}^m$. The surface is defined by the set of equations $\psi_z(\mathbf{x}) = 0$, in which $\{z\} = \{1, \ldots, k\}$ and $k+m=n$. We will work in the following notation: $\mathbf{x}$ represents a dot in $\mathbb{R}^n$ whose coordinates are given by the set $(x^1, x^2, \ldots, x^n)$. All vectorial quantities are represented in the same way.

The Lagrangian can be written as
\begin{equation}  \label{p1}
L(\mathbf{x}, \mathbf{v}) = \frac{1}{2}\ m \mathbf{v}^2 + u^z\psi_z(\mathbf{x})\ ,
\end{equation}
in which $\mathbf{v} = \dot{\mathbf{x}}$ is the velocity of the particle and $u^z$ are Lagrange multipliers. The momenta are given by
\begin{equation*}
\mathbf{p} = \frac{\partial L}{\partial \mathbf{v}} = m\mathbf{v}\ ,\ \ \ \ \ \ \ \ \ \ \ \pi_z = \frac{\partial L}{\partial \dot{u}^z} =0\ .
\end{equation*}

We have by these relations $n$ velocities and $k$ conditions over the momenta. The canonical Hamiltonian function of this system is given by
\begin{equation}
    H_{0} = \mathbf{p}\cdot \mathbf{v} + \pi_{z}\dot{u}^{z} - L = \frac{\mathbf{p}^{2}}{2m} - u^{z}\psi_{z}\ .  \label{p2}
\end{equation}
Therefore, we have the following set of constraints:
\begin{eqnarray}
\phi _{0} &\equiv &\pi _{0}+H_{0}=0\ ,  \label{p3} \\
\phi _{z;1} &\equiv &\pi _{z}=0\ .  \label{p4}
\end{eqnarray}

We should analyze the integrability of this set. The evolution of any dynamical function $F(\mathbf{x}, \mathbf{p}, u^{z}, \pi_{z})$ is given by
\begin{equation*}
   dF = \left\{F,\phi_{0}\right\}dt + \left\{F,\phi_{z;1}\right\}du^{z}\ .
\end{equation*}
Because $\{\phi_{z;1},\phi_{x;1}\}=0$, the condition for $\phi_{z;1}$,
\begin{equation*}
    d\phi_{z;1} = \{\phi_{z;1},\phi_0\}dt + \{\phi_{z;1},\phi_{x;1}\}du^x = 0\ ,
\end{equation*}
gives a secondary set of constraints
\begin{equation*}
\phi_{z;2} \equiv \psi_{z}=0\ ,
\end{equation*}
which are expected since those are the equations of the surface. These constraints must also obey integrability. Hence, we apply
\begin{equation*}
    d\phi_{z;2} = \{\phi_{z;2},\phi_0\}dt + \{\phi_{z;2},\phi_{x;1}\}du^x = 0\ ,
\end{equation*}
which gives a tertiary set:
\begin{equation*}
\phi _{z;3}\equiv \mathbf{p}\cdot \nabla \psi _{z}=0\ .
\end{equation*}
These constraints tells us that the momentum $\mathbf{p}$ is tangent to the surface. And, from these,
\begin{equation*}
    d\phi_{z;3} = \{\phi_{z;3},\phi_0\}dt + \{\phi_{z;3},\phi_{x;1}\}du^x = 0\ ,
\end{equation*}
a quaternary set arise:
\begin{equation*}
\phi _{z;4}\equiv mu^{x}\Delta_{xz}+\mathbf{p}\cdot \nabla\left( \mathbf{p}\cdot \nabla \psi _{z}\right) =0\ ,
\end{equation*}
in which we define the matrix $\Delta_{xz} \equiv \nabla\psi_x\cdot\nabla\psi_z$.

Notice that $\{\phi _{z;4},\phi_0\}\neq0$ and $\{\phi _{z;4},\phi_{x;1}\}\neq0$, so that the integrability of $\phi _{z;4}$ will not give any new constraint, but a total differential equation relating the independent variables. We suppose, therefore, that all constraints are found and the complete set is given by
\begin{eqnarray*}
&&\phi_{0} = p_{0} + H_{0}\ , \\
&&\phi_{z;1} = \pi_{z}\ , \\
&&\phi_{z;2} = \psi_{z}\ , \\
&&\phi_{z;3} = \mathbf{p}\cdot \nabla \psi_{z}\ , \\
&&\phi_{z;4} = m\Delta_{zx}u^{x} + \mathbf{p}\cdot \nabla \left( \mathbf{p} \cdot \nabla \psi_{z}\right)\ .
\end{eqnarray*}

To consider this set in the theory we must build a differential that contains these constraints as generators of the dynamics. However, we do not have independent variables related to all constraints in the system. Therefore, we will expand the parameter space with new arbitrary independent variables $\left(\omega ^{z},\tau ^{z},\theta ^{z}\right)$. Then we define the dynamics of the system by the new fundamental differential
\begin{equation*}
dF=\left\{ F,\phi _{0}\right\} dt+\left\{ F,\phi _{z;1}\right\}du^{z}+\left\{ F,\phi _{z;2}\right\} d\omega ^{z}+\left\{ F,\phi_{z;3}\right\} d\tau ^{z}+\left\{ F,\phi _{z;4}\right\} d\theta ^{z}\ ,
\end{equation*}
as prescribed in section 3.

The above set of HJPDE is not in involution, as we can see with the PB
\begin{equation*}
    \begin{array}{ll}
      \left\{ \phi _{x;1},\phi _{y;2}\right\} = 0 & \left\{ \phi _{x;1},\phi _{y;3}\right\}  = 0 \\
      \left\{ \phi _{x;1},\phi _{y;4}\right\} =-m\Delta _{xy} & \left\{ \phi _{x;2},\phi _{y;3}\right\} =\Delta _{xy} \\
      \left\{ \phi _{x;2},\phi _{y;4}\right\} =\mathbf{p}\cdot \nabla \left(\Delta_{xy}\right) & \left\{ \phi _{x;3},\phi _{y;4}\right\} =\Gamma_{xy}
    \end{array}
\end{equation*}
where
\begin{equation*}
    \Gamma_{xy} \equiv -mu^{z}\nabla \left( \Delta_{yz}\right) \cdot \nabla \psi _{x}+\mathbf{p}\cdot \nabla \left[ \mathbf{p}\cdot \nabla \left( \Delta _{xy}\right) \right] -3\left( \nabla \psi_{x}\right) \cdot \left[ \mathbf{p}\cdot \nabla \left[ \mathbf{p}\cdot\nabla \left( \nabla \psi _{y}\right) \right] \right]\ .
\end{equation*}
We expect, therefore, that all independent variables can be eliminated in function of $t$. Let us reorganize the constraints so that their order in the $M$ matrix becomes $(\phi _{z;1}, \phi _{z;4}, \phi _{z;2}, \phi _{z;3})$. The matrix is given by ($I_x J_y = \{1,4,2,3\}$)
\begin{equation}\label{p5}
    M_{I_xJ_y}=\left(
      \begin{array}{cccc}
        0 & -m\Delta _{xy} & 0 & 0 \\
        m\Delta _{xy} & 0 & -\mathbf{p}\cdot \nabla \left(\Delta_{xy}\right) & - \Gamma_{xy} \\
        0 & \mathbf{p}\cdot \nabla \left(\Delta_{xy}\right) & 0 & \Delta _{xy} \\
        0 & \Gamma_{xy} & -\Delta _{xy} & 0 \\
      \end{array}
    \right)\ .
\end{equation}

Let us suppose that the surface is smooth, of class $C^\infty$. This condition guarantees the existence of the inverse
\begin{equation}\label{p6}
    (M^{-1})^{I_xJ_y} = \left(
      \begin{array}{cccc}
        0 & 1/m\ (\Delta^{-1})^{xy} & \bar{\Gamma}^{xy} & \Sigma^{xy} \\
        -1/m\ (\Delta^{-1})^{xy} & 0 & 0 & 0 \\
        -\bar{\Gamma}^{xy} & 0 & 0 & -(\Delta^{-1})^{xy} \\
        -\Sigma^{xy} & 0 & (\Delta^{-1})^{xy} & 0 \\
      \end{array}
    \right)\ ,
\end{equation}
where
\begin{equation*}
    \bar{\Gamma}^{xy} \equiv (\Delta^{-1})^{xz}\Gamma_{zw}(\Delta^{-1})^{wy}
\end{equation*}
and
\begin{equation*}
    \Sigma^{xy} \equiv \mathbf{p}\cdot \nabla(\Delta^{-1})^{xy}\ .
\end{equation*}
This inverse is unique. Since the matrix is regular all independent variables are indeed dependent of the parameter $t$, as learned with equation (\ref{11}). We are able to define, then, the GB of the system:
\begin{equation}\label{p7}
    \left\{F,G\right\}^* \equiv \left\{F,G\right\} - \left\{F,\phi_{I_x}\right\} (M^{-1})^{I_xJ_y} \left\{\phi_{J_y},G\right\}\ .
\end{equation}
The only non zero fundamental GB are given by
\begin{equation*}
    \{\mathbf{x},\mathbf{p}\}^* = \mathbb{I} - \nabla \psi _{x}(\Delta^{-1})^{xy}\nabla \psi_{y} \equiv \mathcal{P}\ .
\end{equation*}
All others are zero, including $\{u,\pi\}^*$, which is expected since $u^z$ are degenerate variables in the theory. Here we define the projector operator $\mathcal{P}$, whose job is to take vectors in $\mathbb{R}^n$ into vectors on the surface. It is a singular operator, whose base of the null space are the vectors $\nabla\psi_z$. This is expected since normal vectors are projected into null vectors on the surface.

Let us go for the characteristic equations of the reduced system:
\begin{eqnarray}\label{p8}
    d\mathbf{x} &=& \{\mathbf{x},\phi_0\}^*dt = \frac{\mathbf{1}}{m}\ \mathcal{P}\cdot \mathbf{p}\ dt\nonumber \\
    \bar{\mathbf{p}} &=& \mathcal{P}\cdot \mathbf{p}\ ,\\
    \nonumber\\
    \label{p9}
    d\mathbf{p} &=& \{\mathbf{p},\phi_0\}^*dt = \mathcal{P}\cdot u^{z}\nabla\psi_{z}dt = 0\nonumber\\
    \bar{\mathbf{a}} &=& \mathcal{P}\cdot \mathbf{a} = 0\ .
\end{eqnarray}
The first equation gives the projection of the momentum vector over the surface, it is a geometric relation. The second equation is the dynamical equation: it tells us that the acceleration induced on the surface is zero. Of course, we cannot invert the projector, since it is singular, so the equations of motion are still degenerate. Since $\dot{\mathbf{p}} = 0$, the derivative of (\ref{p8}) over time is given by
\begin{equation}\label{p10}
    \ddot{\mathbf{x}} = \frac{d \mathcal{P}}{dt}\cdot \dot{\mathbf{x}}\ .
\end{equation}

Let us suppose the case in which $n=3$ and the surface is just $\mathbb{S}^2$ with equation $\psi = \mathbf{x}^2 - r^2 = 0$. We have $\nabla \psi = 2\mathbf{x}$ and $\Delta = 4\mathbf{x}^2$. In this case $\Delta^{-1} = 1/4r^2$ and
\begin{equation*}
    \mathcal{P} \equiv \mathbb{I} - \frac{\mathbf{x}\otimes\mathbf{x}}{r^2}\ ,
\end{equation*}
where $\otimes$ represents a dyadic product. Using the fact that $\phi_{z;3} = 0$ implies $\mathbf{x}\cdot\dot{\mathbf{x}} = 0$, equation (\ref{p10}) gives
\begin{equation}\label{p11}
    \ddot{\mathbf{x}} + \omega^2\mathbf{x} = 0\ ,
\end{equation}
in which $\omega = v/r$ and $v = |\mathbf{v}|$. The case in which $r=1$ and $m=1$ is an analogous to the linear quantum mechanical $\sigma$-model treated in \cite{rothe}.

\section{The multi-dimensional rotator}

Let us work with the Lagrangian \cite{rothe}
\begin{equation}
L=\frac{1}{2}\ \mathbf{v}^{2}+u\mathbf{x}\cdot \mathbf{v}\ ,  \label{r1}
\end{equation}
where $\mathbf{v}=\dot{\mathbf{x}}$. This system constitutes a more general case than the previous example. The conjugate momenta are given by
\begin{eqnarray*}
\mathbf{p} &=&\mathbf{v}+u\mathbf{x}\ , \\
\pi &=&0\ .
\end{eqnarray*}

The canonical Hamiltonian gives
\begin{equation}
H_{0}=\frac{1}{2}\left[ \mathbf{p}-u\mathbf{x}\right] ^{2}=\frac{1}{2}\ \bar{\mathbf{p}}^{2}\ ,  \label{r2}
\end{equation}
in which we define $\bar{\mathbf{p}}\equiv \mathbf{p}-u\mathbf{x}$. Therefore, the system has the constraints
\begin{eqnarray}
\phi _{0} &=&p_{0}+H_{0}\ ,  \label{r3} \\
\phi _{1} &=&\pi \ .  \label{r4}
\end{eqnarray}

The fundamental differential is, in this case,
\begin{equation*}
dF = \{F,\phi_0\}dt + \{F,\phi_1\}du\ .
\end{equation*}
The integrability condition for $\phi_1$ gives

\begin{equation*}
d\phi _{1}=\{\phi _{1},\phi _{0}\}dt+\{\phi _{1},\phi _{1}\}du=\bar{\mathbf{p%
}}\cdot \mathbf{x}dt\ ,
\end{equation*}%
which results in a new constraint
\begin{equation}
\phi _{2}=\bar{\mathbf{p}}\cdot \mathbf{x}=\mathbf{p}\cdot \mathbf{x}-u%
\mathbf{x}^{2}=0\ .  \label{r5}
\end{equation}%
The integrability of $\phi _{2}$ gives no new constraint. The system is not
in involution, giving a regular $M$ matrix and its inverse
\begin{equation*}
M=\left(
\begin{array}{cc}
0 & \mathbf{x}^{2} \\
-\mathbf{x}^{2} & 0%
\end{array}%
\right) \ ,\ \ \ \ \ \ \ \ \ \ \ \ \ M^{-1}=\left(
\begin{array}{cc}
0 & -1/\mathbf{x}^{2} \\
1/\mathbf{x}^{2} & 0%
\end{array}%
\right) \ .
\end{equation*}

The GB of this problem, that will eliminate the independent variables in
function of $t$, is given by
\begin{equation*}
\left\{ F,G\right\} ^{\ast }=\left\{ F,G\right\} -\left\{ F,\phi
_{1}\right\} (M^{-1})^{12}\left\{ \phi _{2},G\right\} -\left\{ F,\phi
_{2}\right\} (M^{-1})^{21}\left\{ \phi _{1},G\right\} \ ,
\end{equation*}%
which gives the following GB:%
\begin{eqnarray*}
\{\mathbf{x},\mathbf{\bar{p}}\}^{\ast } &=&\mathbb{I}-\frac{\mathbf{x}%
\otimes \mathbf{x}}{\mathbf{x}^{2}}\equiv \mathcal{P}\ , \\
\{\mathbf{x},u\}^{\ast } &=&\frac{\mathbf{x}}{\mathbf{x}^{2}}\ ,\ \ \ \ \ \ \ \ \ \ \{u,\mathbf{\bar{p}}\}^{\ast }\ =\ \frac{\bar{\mathbf{p}}}{\mathbf{x}^{2}}\ .
\end{eqnarray*}

The characteristics equations are given by
\begin{eqnarray*}
d\mathbf{x} &=&\{\mathbf{x},\phi _{0}\}^{\ast }dt=\mathcal{P}\cdot \bar{%
\mathbf{p}}\ dt\ , \\
d\bar{\mathbf{p}} &=&\{\bar{\mathbf{p}},\phi _{0}\}^{\ast }dt=0\ , \\
du &=&\{u,\phi _{0}\}^{\ast }dt=\frac{\bar{\mathbf{p}}^{2}}{\mathbf{x}^{2}}\
dt\ .
\end{eqnarray*}%
Hence, we have the equations%
\begin{eqnarray}
\mathbf{v} &=&\mathcal{P}\cdot \bar{\mathbf{p}}\ ,  \label{r6} \\
\dot{\bar{\mathbf{p}}} &=&0\ .  \label{r7}
\end{eqnarray}

Derivation of the first equation over time and the use of the second
equation gives%
\begin{equation*}
\ddot{\mathbf{x}}=\dot{\mathcal{P}}\cdot \bar{\mathbf{p}}=-\frac{\mathbf{v}%
\cdot \bar{\mathbf{p}}}{\mathbf{x}^{2}}\ \mathbf{x}\ ,
\end{equation*}%
where we used the constraint $\bar{\mathbf{p}}\cdot \mathbf{x}=\mathbf{0}$,
which also implies, from (\ref{r6}), $\mathbf{v}=\mathbf{\bar{p}}$. We have,
therefore, the nonlinear second order ODE%
\begin{equation}
\ddot{\mathbf{x}}\cdot \mathbf{x}+\dot{\mathbf{x}}^{2}=0\ .  \label{r8}
\end{equation}%
This equation can also be written by%
\begin{equation*}
\frac{d^{2}\mathbf{x}^{2}}{dt^{2}}=0\ .
\end{equation*}

Notice that the equation of motion (\ref{r8}) is obeyed by any system in
which $\mathbf{x}\cdot \dot{\mathbf{x}}$ is a constant. The most simple
particular solution is given by $|\mathbf{x|}=\sqrt{\alpha t+\beta }$.
However, considering a system in which $\mathbf{x}^{2}=r^{2}$ is fixed and $|%
\mathbf{v|}$ is constant, we can build a solution%
\begin{equation*}
\mathbf{x}=\mathbf{a}\sin \omega (t)+\mathbf{b}\cos \omega (t)
\end{equation*}%
where $\mathbf{a}$ and $\mathbf{b}$ are constant vectors and%
\begin{equation*}
\omega (t)=\frac{1}{r}\left( |\mathbf{v}|t+\beta \right) ,
\end{equation*}%
where $\beta $ is a parameter. This solution is of the same class of the one
found in \cite{rothe} for the n dimensional case.

\section{Landau model}

Consider the Lagrangian \cite{rothe}
\begin{equation*}
    L = \frac{1}{2}\left(mv_iv_i + B \epsilon_{ij}x_iv_j - kx_ix_i\right)\ \ \ \ \ \ \ \ \ \ \ \{i\} = \{1,2\}\ ,
\end{equation*}
which represents a charged particle in a plane with a transversal magnetic field $B$ and a harmonic potential. In the limit $m=0$ this system is described by
\begin{equation}\label{l1}
    L = \frac{1}{2}\left(B \epsilon_{ij}x_iv_j - kx_ix_i\right)\ .
\end{equation}
The conjugate momenta give the relations
\begin{equation*}
    p_i = \frac{1}{2}\ B \epsilon_{ij}x_j\ .
\end{equation*}

Let us write the canonical Hamiltonian:
\begin{equation}\label{l2}
    H_0 = \frac{1}{2}\ kx_ix_i\ ,
\end{equation}
that gives us the system of HJPDE
\begin{eqnarray}
  \label{l3}
  \phi_0 &=& p_0 + H_0 = 0\ ,\\
  \label{l4}
  \phi_i &=& p_i + \frac{1}{2}\ B \epsilon_{ij}x_j = 0\ .
\end{eqnarray}

This problem is completely constrained and all $x$ variables are parameters in the theory. The fundamental differential is
\begin{equation*}
    dF = \{F,\phi_0\}dt + \{F,\phi_i\}dx^i\ .
\end{equation*}
Since $\{\phi_i, \phi_j\} = B \epsilon_{ij}$, the constraints are not involutive and the integrability conditions for $\phi_i$ will not give any constraint, but a total differential equation relating the variables $x$ and $t$. We have the $M$ matrix and the inverse
\begin{equation*}
    M_{ij} = B \epsilon_{ij}\ \ \ \ \ \ \ \ \ \ \ \ M^{-1}_{ij} = - \frac{1}{B}\ \epsilon_{ij}\ .
\end{equation*}
The GB are defined by
\begin{equation}
    \label{l5}\{F,G\}^* \equiv \{F,G\} + \frac{1}{B}\ \{F,\phi_i\}\epsilon_{ij}\{\phi_j,G\}\ .
\end{equation}
Therefore,
\begin{equation*}
    \{x_i, x_j\}^* = - \frac{1}{B}\ \epsilon_{ij}\ ,\ \ \ \ \ \ \{x_i, p_j\}^* = \frac{1}{2}\ \delta_{ij}\ ,\ \ \ \ \ \ \{p_i, p_j\}^* = - \frac{1}{4}\ B \epsilon_{ij}\ .
\end{equation*}

The equations of motion are given by
\begin{eqnarray*}
  dx_i &=& \{x_i, \phi_0\}^*dt =  - \frac{1}{B}\ k\epsilon_{ij}x_jdt\ ,\\
  dp_i &=& \{p_i, \phi_0\}^*dt = - \frac{1}{2}\ k x_idt \ .
\end{eqnarray*}
The result is
\begin{equation}\label{l6}
    \ddot{x}_i + \frac{k^2}{B^2}\ x_i = 0\ ,
\end{equation}
in full agreement with \cite{rothe}.

\section{The Güler's example}

Consider the following Lagrangian function \cite{guler1}
\begin{equation}\label{g1}
    L = \frac{1}{2}\ \dot{q}_1^2 - \frac{1}{4}\ \left(\dot{q}_2 - \dot{q}_3\right)^2 + b\dot{q}_2 - c\ ,
\end{equation}
in which $b$ and $c$ are functions of $q_1$, $q_2$, $q_3$ and $t$. The momenta are given by
\begin{equation*}
  p_1 = \dot{q}_1 \ , \ \ \ \ \ \ \ \ p_2 = - \frac{1}{2}\left(\dot{q}_2 - \dot{q}_3\right) + b \ , \ \ \ \ \ \ \ p_3 = \frac{1}{2}\left(\dot{q}_2 - \dot{q}_3\right)\ .
\end{equation*}

Therefore, there is a constraint relating the momenta $p_2$ and $p_3$:
\begin{equation*}
    p_2 + p_3 - b = 0\ .
\end{equation*}
We will choose to write the Hamiltonian in terms of $p_1$ and $p_2$, which implies to choose $q_1$ and $q_2$ as dependent variables, while $q_3$ ant $t$ are the independent ones. We have, then,
\begin{equation}\label{g2}
    H_0 = \frac{1}{2}\ p_1^2 - \left(b - p_2\right)^2 + c\ ,
\end{equation}
which gives us the set
\begin{eqnarray}
  \label{g3}
  \phi_0 &\equiv& p_0 + H_0\ , \\
  \label{g4}
  \phi_1 &\equiv& p_2 + p_3 - b = 0\ .
\end{eqnarray}

The fundamental differential is given by
\begin{equation*}
    dF = \{F, \phi_0\}dt + \{F, \phi_1\}dq_3\ .
\end{equation*}
With this differential we test the integrability of $\phi_1$, which gives a new constraint
\begin{equation*}
    2\left(b - p_2\right)\frac{\partial b}{\partial q_3} - \frac{\partial b}{\partial t} - p_1\frac{\partial b}{\partial q_1} - \frac{\partial c}{\partial q_2} - \frac{\partial c}{\partial q_3} = 0\ .
\end{equation*}
With $b = q_1 + q_3$ and $c = q_1 + q_2 + q_3^2$ this constraint can be written as
\begin{equation}\label{g5}
    \phi_2 \equiv 2p_2 + p_1 - 2q_1 + 1 = 0\ .
\end{equation}
We have $\{\phi_2, \phi_1\} = 1$ and no new constraints. The $M$ matrix and its inverse follows:
\begin{equation*}
    M = \left(
          \begin{array}{cc}
            0 & -1 \\
            1 & 0 \\
          \end{array}
        \right)\ , \ \ \ \ \ \ \ \ \ \ M^{-1} =
    \left(
      \begin{array}{cc}
        0 & 1 \\
        -1 & 0 \\
      \end{array}
    \right)\ .
\end{equation*}

The GB are defined by
\begin{equation}
    \label{g6}
    \{F,G\}^* \equiv \{F,G\} - \{F,\phi_1\}\{\phi_2,G\} + \{F,\phi_2\}\{\phi_1,G\}
\end{equation}
and gives the following non zero fundamental GB:
\begin{eqnarray*}
  &&\{q_1,q_2\}^* = \{q_1,q_3\}^* = \{q_1,p_3\}^* = -1\\
  &&\{q_2,q_3\}^* =  \{q_2,p_3\}^* = \{p_1,p_3\}^* = -2\\
  &&\{q_2,p_2\}^* = \{q_3,p_3\}^* = 1\\
  &&\{q_3,p_1\}^* = 2\ .
\end{eqnarray*}

Those give us the equations
\begin{eqnarray*}
  dq_1 &=& \{q_1,\phi_0\}^*dt = \left(2q_1 - 2p_2 - 1\right)dt \\
  dq_2 &=& \{q_2,\phi_0\}^*dt = \left(4q_1 - 4p_2 + 1\right)dt \\
  dq_3 &=& \{q_3,\phi_0\}^*dt = \left(2p_1 + 2p_2 - 2q_1 - 2q_3 + 3\right)dt \\
  \\
  dp_1 &=& \{p_1,\phi_0\}^*dt = \left(4q_1 + 2q_3 - 4p_2\right) dt \\
  dp_2 &=& \{p_2,\phi_0\}^*dt = -1 dt \\
  dp_3 &=& \{p_3,\phi_0\}^*dt = \left(2p_1 - 2q_3 + 3\right) dt\ ,
\end{eqnarray*}
which can be written, using the constraints, by
\begin{equation}
  \label{g7}
  \ddot{q}_1 = 2\left(\dot{q}_1 + 1\right)\ , \ \ \ \ \ \ \ \ \ \ \ddot{q}_2 = 4\left(\dot{q}_1 + 1\right)\ , \ \ \ \ \ \ \ \ \ \ \ddot{q}_3 = 4\left(q_3 - \frac{1}{2}\right) \ .
  \end{equation}
As we can see, the solutions in \cite{guler1} satisfies these equations.

\section{Final Comments}

The appearance of non-involutive constraints violates the Frobenius' integrability condition, by which the constraints related to a set of LI Hamiltonian vector fields must close a Lie algebra with the Poisson brackets. In this work we have shown that the imposition of the integrability conditions $d\phi_\alpha = 0$ provides a self consistent way to treat non-involutive (second class) constrained systems.

It must be stressed, however, that the procedure do not change the Frobenius' theorem: if a system is in agreement with (\ref{08}) it is also in agreement with (\ref{07}). We saw that the imposition of (\ref{08}) can reveal new constraints or show that the system, or part of it, is integrable, cases that are covered by (\ref{07}), but it can also reveal linear dependence in the vector fields tangent to the surfaces, which implies elimination of independent variables related to non-involutive constraints.

The process of elimination shows that it is always possible to redefine the dynamics with the introduction of Generalized brackets. The GB actually defines a symplectic structure in a reduced phase space. Systems that presents only non-involutive constraints ($M$ regular case) are completely reduced and the evolution of any observable is generated by the canonical Hamiltonian with the GB, as shows equation (\ref{13}). Therefore, non-involutive constraints are not dynamical generators in the reduced phase space. This is reflex of the fact that the constraints that were not in involution with the PB are in involution with the GB now.

However, if the system presents secondary constraints a previous step is needed. Since there are no independent variables related to constraints that come from the integrability conditions, we may expand the parameter space to include new parameters, one for each of those constraints. The secondary constraints are, then, assumed to be generators of infinitesimal transformations\footnote{This statement can always be made, but we stress that secondary constraints may not be dynamical generators in the reduced phase space. This is actually the case with non-involutive constraints. For involutive set of constraint the validity and consequences of this statement are still under investigation.} in the phase space, in equality with the primary constraints, as shown in the evolution equation (\ref{09}).

If the system has only a subset of non-involutive constraints we have the case in which $M$ is singular. This subset of $r$ constraints is computed in the GB of the system. However, integrability conditions for the other $k-r$ constraints are given by equations (\ref{19}), which can only result in new relations between the variables (in this case new constraints), or in identical $0=0$ relations. The primary case requires the insertion of these new constraints in the system. However, this case will not occur if all secondary constraints are already previously found. The second case implies that the remaining constraints must be in involution with the GB and the system whose evolution differential is given by (\ref{17}) is already completely integrable.

\section*{Acknowledgments}

MCB was supported by CAPES. CEV was supported by CNPq. BMP was partially supported by CNPq. The authors would like to thank P. J. Pompeia for the critical reading and suggestions.

\end{document}